\documentclass[12pt]{article}
\usepackage{amsmath,epsfig,amssymb,eufrak}
\usepackage{amscd,latexsym}
\usepackage{color}

\date{}

\def\xlf{\raisebox{+0.2em}{\color{red}\boldmath{$\chi$}}\hspace{-0.2ex}\raisebox{-0.2em}{\color{green}L}
\hspace{-1.5ex}\raisebox{+0.14em}{\color{blue}F}\hspace{2mm}}

\def\lsi{\raise0.3ex\hbox{$<$\kern-0.75em\raise-1.1ex\hbox{$\sim$}}}
\def\gsi{\raise0.3ex\hbox{$>$\kern-0.75em\raise-1.1ex\hbox{$\sim$}}}

\input macros.sty      
\input eqalign.sty     

\begin{document}

\begin{titlepage}

\title{
  {\vspace{-0cm} \normalsize
  \hfill \parbox{40mm}{DESY/05-131\\
                       SFB/CPP-05-31\\
                       LTH657\\
                       July 2005}}\\[10mm]
Flavour Breaking Effects \\ of Wilson twisted mass fermions}  
\author{ K.~Jansen$^{\, 1}$, C. McNeile$^{\, 3}$, C. Michael$^{\, 3}$,  
K. Nagai$^{\, 1}$, M.~Papinutto$^{\, 1}$, \\ J. Pickavance$^{\, 3}$,  
A.~Shindler$^{\, 1}$, C.~Urbach$^{\, 1,2}$ and I.~Wetzorke$^{\, 1}$ 
\\
\\
   {\bf \xlf Collaboration}\\
\\
{\small $^{1}$  John von Neumann-Institut f\"ur Computing NIC,} \\
{\small         Platanenallee 6, D-15738 Zeuthen, Germany} \\ \ \\
{\small $^{2}$  Institut f\"{u}r Theoretische Physik, Freie Universit\"{a}t Berlin,} \\
{\small Arnimallee 14, D-14195 Berlin, Germany}\\ \ \\
{\small $^{3}$  Theoretical Physics Division, Dept of Mathematical Sciences,} \\  
{\small University of Liverpool, Liverpool L69 3BX, UK.}
}

\maketitle

\begin{abstract}
 We study the flavour breaking effects appearing in the Wilson twisted
mass  formulation of lattice QCD. In this quenched study, we focus on 
the mass splitting between
 the neutral and the charged pion, determining the neutral pion mass
with a stochastic noise method to  evaluate the disconnected
contributions. We find that these disconnected contributions are
significant. Using the Osterwalder-Seiler  interpretation of 
the {\em connected} piece of  the neutral
pion correlator, we compute the corresponding neutral pion mass to study
with more precision the scaling behaviour of the  mass splitting. 
  \vspace{0.75cm}
\noindent
\end{abstract}

\end{titlepage}

\section{Introduction}

Wilson twisted mass fermions \cite{Frezzotti:2000nk,Frezzotti:2003ni} 
have by now been employed extensively in  numerical simulations of
lattice QCD, both quenched 
\cite{Jansen:2003ir,Bietenholz:2004wv,Jansen:2005gf,Jansen:2005kk,Abdel-Rehim:2004gx,Abdel-Rehim:2005gz}
and unquenched 
\cite{Farchioni:2004us,Farchioni:2004ma,Farchioni:2004fs,Farchioni:2005tu}
with  very promising results: In the quenched case, the anticipated 
$O(a)$-improvement for physical observables at full twist 
\cite{Frezzotti:2003ni} has been demonstrated and, employing
a definition of the critical mass derived from the vanishing of the 
PCAC quark mass  
\cite{Sharpe:2004ny,Aoki:2004ta,Frezzotti:2005gi}, it has also been
demonstrated numerically \cite{Jansen:2005gf,Jansen:2005kk,Abdel-Rehim:2005gz}
and analytically \cite{Frezzotti:2005gi} that $O(a^2)$ effects are small 
for physical quantities even at low values of the pion mass of about $270$ MeV.
These results are very encouraging, in particular if one thinks of eventual
dynamical simulations at full twist which we have just started. 

One important property of the twisted mass approach is that at  a
non-vanishing value of the lattice spacing $a$, flavour symmetry is explicitly 
broken. This  manifests itself, e.g. in the mass splitting, as a
non-vanishing difference of the charged $m_{\pi^+}$ and the neutral
$m_{\pi^0}$ pion masses.  Of course, this phenomenon is a pure cut-off
effect which is expected to vanish, at full twist, quadratically with
the lattice spacing  when the continuum  limit is approached.  We will
concentrate in this work on the difference  $m_{\pi^0}-m_{\pi^+}$ to
study the flavour breaking effects and their lattice spacing dependence.
 We will compute the neutral pion mass from the connected and 
disconnected pieces of the corresponding correlation function  using  
the stochastic noise source method of 
refs.~\cite{Foster:1998vw,McNeile:2000hf} for the disconnected diagrams.
 In quenched studies, the disconnected contributions  to flavour singlet
correlators are known to be pathological - having a double pole 
structure. 
Note, however, that for the twisted mass situation 
even in the quenched approximation in the continuum limit
the full neutral pion correlation function (connected plus disconnected 
parts) is a 
well-defined correlator since at $a=0$ the flavour symmtry is restored and 
the twisting is just a formal 
rotation that leaves the theory invariant. Therefore the continuum 
limit of the neutral pion correlator evaluated at a non-vanishing value 
of the lattice spacing can be taken.
Additionally, we consider another definition of the neutral pion mass
from the connected piece of the twisted mass neutral pion 
correlator alone,  using the  Osterwalder-Seiler
interpretation \cite{Osterwalder:1977pc,Frezzotti:2004wz}  of 
this correlator.
This last procedure is only valid in the  quenched approximation
and will  mainly provide a more accurate estimate of the scaling
properties  of the flavour breaking cut-off  effects, checking the
strength of the $O(a^2)$ cut-off effects.

We will work with two  definitions for the critical quark
mass. The first is the point where  the pion mass vanishes, the second,
where the  PCAC quark mass vanishes. In the following we will refer to
the first situation as the ``pion definition'' and to the second
situation  as the ``PCAC definition'' of the critical point. Both
definitions lead to an $O(a)$-improvement, but  they can lead to very
different $O(a^2)$ effects \cite{Frezzotti:2005gi}, 
in particular at  small pion masses.  In the
language of twisted mass lattice QCD,  tuning to the critical quark mass
corresponds to working at full twist. In 
refs.~\cite{Jansen:2005gf,Jansen:2005kk} 
we presented results from both definitions of the  critical bare quark 
mass and demonstrated that the PCAC definition shows indeed
small scaling violations even  when the pion mass is taken to be as low
as $270$ MeV.  Here we are interested to see, whether also for the pion
mass  difference a similar situation occurs.

While it is the primary goal of this paper to study the strength of
flavour breaking effects in Wilson twisted mass QCD, another interesting
question is the evaluation of the disconnected diagrams themselves
within this approach.  Such disconnected diagrams are needed in many
physical problems such as  the computation of the pseudoscalar and
scalar  flavour singlet masses,  the decay of the $\rho$-meson, the
phenomenon of string breaking and  contributions to the vacuum
polarisation tensor.    
In addition, if one thinks of simulations with mixed actions, it 
becomes again important to know, how reliably the disconnected
diagrams can be computed.
 It is hence of an essential interest to know whether the disconnected 
pieces can be 
computed with a reasonable number, say $O(100-500)$, of independent
gauge field  configurations and $O(10-50)$ stochastic noise vectors.  

\section{Wilson twisted mass fermions}

The Wilson twisted mass action in the so-called twisted basis 
can be written \cite{Frezzotti:2003ni} as 
\begin{equation}
  \label{tmaction}
  S[U,\chi,\bar\chi] = a^4 \sum_x \bar\chi(x) ( D_W + m_0 + i \mu
\gamma_5\tau_3 ) \chi(x)\equiv \bar\chi D_\mathrm{tm}\chi\; ,
\end{equation}
where the Wilson-Dirac operator $D_{\rm W}$ is given by 
\begin{equation}
  D_{\rm W} = \sum_{\mu=0}^3 \frac{1}{2} 
  [ \gamma_\mu(\nabla_\mu^* + \nabla_\mu) - a \nabla_\mu^*\nabla_\mu]
  \label{Dw}
\end{equation}
and $\nabla_\mu$ and $\nabla_\mu^*$ denote the usual forward
and backward derivatives and the Wilson parameter $r$ was set to $1$. 

The situation of full twist and hence automatic $O(a)$ improvement
arises when $m_0$ in eq.~(\ref{tmaction}) is tuned towards a critical
bare quark mass $m_{\mathrm{c}}$. We use for our simulations the
hopping representation of the Wilson-Dirac operator with
$\kappa=(2am_0+8)^{-1}$ and hence this critical quark mass corresponds
to a critical hopping parameter 
$\kappa_{\mathrm{c}}=(2am_{\mathrm{c}}+8)^{-1}$.
As remarked in the introduction, in this paper we will use the 
pion and the PCAC definitions of $\kappa_\mathrm{c}$ to realize 
full twist.

We consider the local bilinears
$P^\pm =\bar\chi\gamma_5\frac{\tau^\pm}{2}\chi$ and
$P^0 =\bar\chi\chi$ (corresponding to the scalar operator in the 
twisted basis), 
where $\chi$ denotes a mass-degenerate doublet of up and down quarks. 
From these operators we 
extract the charged $m_{\pi^{+}}$ and neutral 
$m_{\pi^{0}}$ pseudoscalar masses 
from the correlation functions:
\begin{equation}
  \begin{split}
    C_{\pi^{+}} (x_0) &= a^3\sum_{\mathbf x} \langle 
                         \left[P^+(x)P^-(0)\right]_\mathrm{con}\rangle\ , \\
    C_{\pi^{0}} (x_0) &= a^3\sum_{\mathbf x} \langle  
                         \left[P^0(x)P^0(0)\right]_\mathrm{con} 
                      +   \left[P^0(x)P^0(0)\right]_\mathrm{disc}\rangle 
  \end{split}
\label{correlations}
\end{equation}
\noindent where  
\begin{equation}
\left[P^0(x)P^0(0)\right]_\mathrm{disc} = 
\left[ \mathrm{tr} D^{-1}_\mathrm{tm}\right](x) 
               \left[ \mathrm{tr} D^{-1}_\mathrm{tm}\right](0)
\label{disconnected}
\end{equation}
with the vacuum contribution to $\mathrm{tr} D^{-1}_\mathrm{tm}$ being
subtracted and the trace taken over colour and Dirac indices.
We indicate in eqs.~(\ref{correlations}) 
the connected (con) and the disconnected (disc) pieces of the
correlation function and denote by $\left[\;\right]$ the fermionic contractions
only.

We also checked the operator 
$A_0^3=\bar\chi\gamma_0\gamma_5\frac{\tau^3}{2}\chi$
to compute the neutral pion correlation function which has the advantage 
to not develop a vacuum contribution. However, we found that this operator 
is very noisy for the disconnected contribution 
and could not be used to extract a signal. 
We see two possible explanations for this behaviour. The first is that the 
operator itself is proportional to the pion mass. 
Hence, the signal becomes small at low values of the pion 
mass while the error remains roughly constant. 
The other reason can be that in the free theory the correlation function
is identically zero since from the fermionic contractions there remains only
a $\mathrm{tr}\gamma_0$. Consequently, the correlation function starts 
with a $a^2g_0^2$ behaviour, leading again to a possible small signal. 
We cannot disentangle what is the major effect for the observed large
fluctuations in the $A_0^3$ correlation function, or, whether it is a 
mixture of both. Anyway, we conclude that the $A_0^3$ operator is, 
from a practical point of view, not a suitable operator to study the 
neutral pion correlation function. 

\section{Results}

In this section we present our simulation results. We use a 
stochastic source method, for which we only employ the PCAC definition for
the critical point. We will also use the connected part of the neutral
pion correlation function which we interprete with the help of the 
Osterwalder-Seiler action. For this part, we use both, pion and 
PCAC definitions
of the critical point. 

\subsection{Stochastic source method}

The correlation function for the neutral pion in
eq.~(\ref{correlations}) has a connected piece as well as a disconnected
piece. While the connected piece can be computed by standard methods
using local or smeared sources,  we evaluate  the disconnected piece by
stochastic source   methods. We follow here the techniques used in 
refs.~\cite{McNeile:2000xx,McNeile:2001cr}, in particular  we used a
stochastic source method with variance reduction as described in  the
appendix of ref.~\cite{McNeile:2000xx}. Given a number of gauge field
configurations, see  table~\ref{table:datastochastic} for our
statistics, we performed the evaluation of the disconnected piece of the
  correlation function for a number $N_\Phi$ stochastic noise vectors
$\Phi$. We found that for $N_\Phi=24$ the contribution of the 
stochastic sources to the total error  is less or comparable to the
error from the gauge field average.
 Hence, for the stochastic  estimate of the disconnected pieces of the
correlation function  $ C_{\pi^{0}} (x_0)$ in eq.~(\ref{correlations})
we always used  $N_\Phi=24$. We measure the disconnected contributions
for both  local and fuzzed (non-local) \cite{Lacock:1994qx}
operators which gave consistent results. Let us mention that we
used as a check the stochastic source method also for the connected
piece of the correlation functions  in eq.~(\ref{correlations}) and
found full agreement  with the computation by standard methods. 
 In table~\ref{table:datastochastic} we give the parameter values  for
$\beta$, $\mu$ and  $\kappa_{\mathrm{c}}$ that we have used for the
analysis using the stochastic source  method. Note that here we only
employ the PCAC definition of $\kappa_{\mathrm{c}}$ to realize full twist. 
Table~\ref{table:datastochastic} contains our results for the  neutral
pion mass $m_{\pi^0}$ and the mass splitting  $m_{\pi^0}-m_{\pi^+}$. 

\begin{table}[!t]
\begin{center}
\begin{tabular}{|c||c|c|c|c|c|}
\hline\hline
 N$_\mathrm{meas}$ &  $\beta$ &  $\kappa_{\mathrm{c}}$ & $\mu a$  & 
 $m_{\pi^0}a$(tot)& $m_{\pi^0}a-m_{\pi^+}a$\\
\hline\hline
 400 & 6.0  & 0.157409 & 0.0038 & 0.19(2) & 0.07(2)\\
 400 & 6.0  & 0.157409 & 0.0076 & 0.19(2) & 0.02(2)\\
 400 & 6.0  & 0.157409 & 0.0109 & 0.23(2) & 0.02(2)\\
\hline                                                   
 100 & 5.85 & 0.162379 & 0.0050 & 0.28(5) & 0.12(5)\\
 100 & 5.85 & 0.162379 & 0.0100 & 0.25(5) & 0.02(5)\\
 100 & 5.85 & 0.162379 & 0.0144 & 0.28(6) & 0.00(6)\\
\hline\hline
\end{tabular}
\end{center}
 \caption{\it The number of gauge field configurations N$_\mathrm{meas}$ 
as well the
values of $\beta$,  $\kappa_{\mathrm{c}}$ and  $\mu$ we have used for
the  stochastic source method. We give the values of the neutral pion
mass $m_{\pi^0}$  evaluated from fits to the full correlation function 
$C_{\pi^{0}} (x_0)$ in eq.~(\ref{correlations}) and the mass splitting  
$m_{\pi^0}-m_{\pi^+}$.
Data for the charged pion mass $m_{\pi^+}$ and the neutral pion mass 
computed from the connected piece of the correlation function can be 
found in tables~\ref{table:piplus} and \ref{table:pineutral}. 
The computations used
$N_\Phi=24$ stochastic noise sources to determine the disconnected
pieces of the correlation functions.
 }
\label{table:datastochastic}
\end{table}

\begin{figure}[htb]
\begin{center}
\vspace*{-4cm}
\epsfig{file=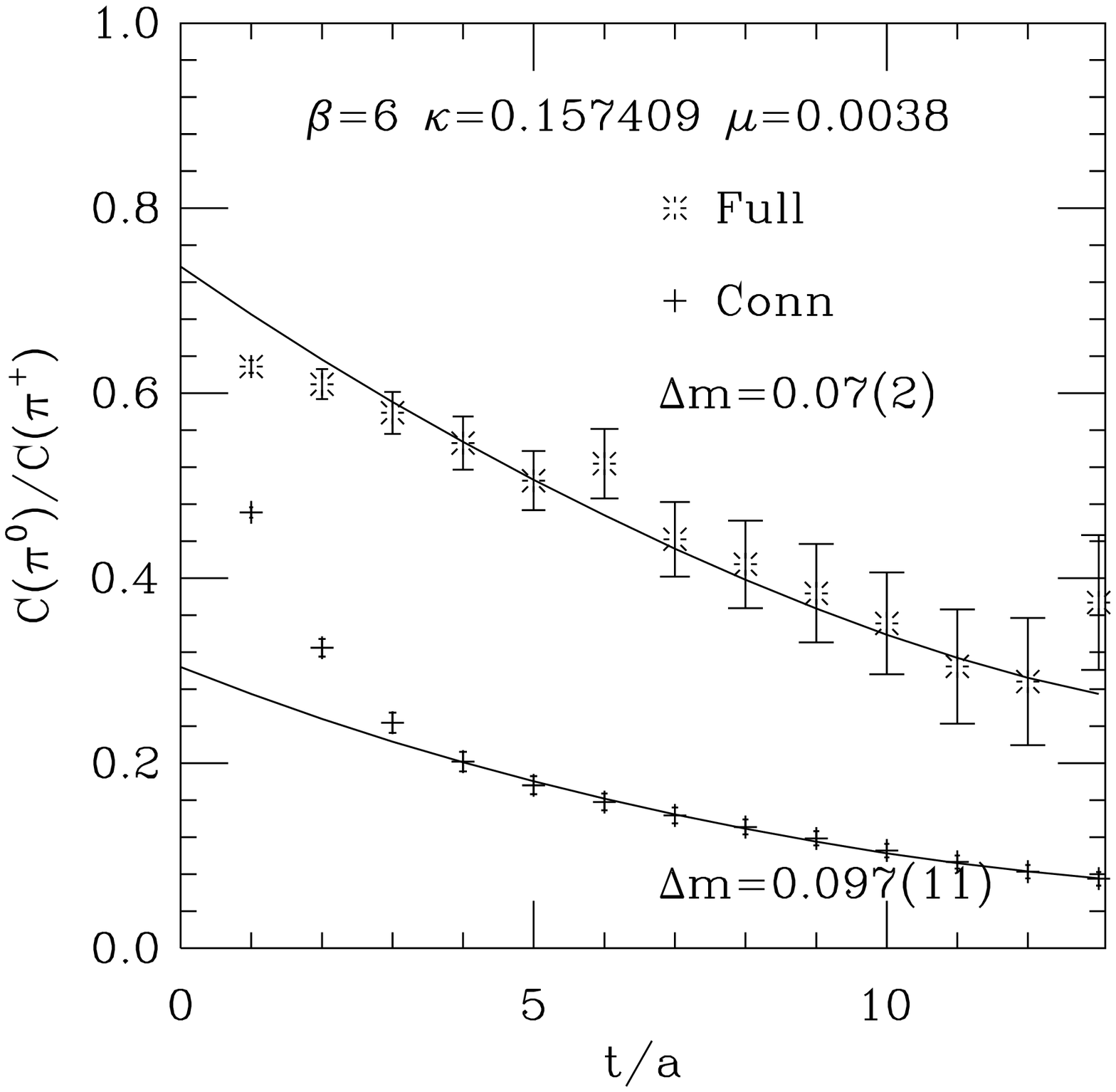,width=18.0cm}
\end{center}
 \caption{Correlation function ratios (for local operators only) of the
neutral to the charged  pion correlation functions for both the full
correlation functions  in eq.~(\ref{correlations}) and taking only the
connected part of the  neutral pion correlation function. The solid
lines are the corresponding ratios from the fits to these correlation
functions. We also indicate the  values of mass splitting $\Delta
m=m_{\pi^0}-m_{\pi^+} >0 $, see also  table~\ref{table:datastochastic}.
 \label{fig:ratiocorr}
 }
\end{figure}

In fig.~\ref{fig:ratiocorr} we show, as an example, the ratio of the 
correlation functions of the neutral and the charged pion at $\beta=6.0$
and $\mu a=0.0038$.  We show the ratio of the correlation functions for
the connected  part only and for the full correlation function including
the disconnected piece.  Clearly, the ratio of these correlation
functions is not constant and  does not assume a value of one as 
would be the case for a situation  where no flavour violation
appears.  In fact, the  deviations from one of the correlation function
ratio appears to be  rather large indicating strong flavour breaking
effects. Note,  however, that these effects are substantially larger
when only the  connected pieces are considered and the addition of the 
disconnected piece reduces the flavour symmetry violation significantly.
 We also remark that the ratio of pion correlators is closer to 1.0 at
$\beta=6.0$ compared to $\beta=5.85$. This indicates 
that the flavour breaking effects become small when the continuum limit 
is approached as can also be seen from the comparison of  
$\Delta m=m_{\pi^0}-m_{\pi^+} $ in table~\ref{table:datastochastic}. We will
come back to this point in the next section.  The time dependence of the
ratio in the graph reflects the fact that the  mass splitting  $\Delta
m$ is non-vanishing and positive. The solid lines in 
fig.~\ref{fig:ratiocorr} are the ratio of the fit functions  obtained by
separate fits to the neutral and charged pion correlation  functions.  

We also considered the flavour splitting between the charged and neutral
states of the  scalar meson with $I=1$, the $a_0$. Consider the
correlation function of the bilinear operator  $a_0^0
=\bar\chi\gamma_5\chi$ and the mass splitting with the corresponding
charged state $a_0^{+}$ where the correlation function for $a_0^{0}$
again has a disconnected piece. As for the $\pi$, we find that the 
disconnected contribution to $a_0^0$ is such as to reduce the flavour 
violation from that observed with the connected piece alone.  However, 
both  correlation functions (for  $a_0^0$ and $a_0^+$) become  negative 
as $t$ increases, demonstrating the failure of the quenched
approximation. We finally remark that also for the $a_0^{+}$, $a_0^{0}$
system the flavour  breaking effects are reduced when moving from
$\beta=5.85$ to  $\beta=6.0$ indicating that this effect will disappear in
the continuum limit. 

\subsection{Osterwalder-Seiler interpretation} 

The calculation of the disconnected piece for the neutral 
pion correlation function leads to a rather large error for the 
neutral pion mass. 
One may wonder, whether the neutral pion mass could not be extracted
from the connected piece of this correlation function alone. 
However, 
in principle neglecting the disconnected diagram may lead to a correlator
that does not have an interpretation in terms of a local operator and so does
not have an
interpretation in terms of the transfer matrix.
One way out of this problem is to 
use the Osterwalder-Seiler (OS) action to interprete the 
connected piece of the twisted mass neutral pion correlation function. 
As 
we will see, the connected piece can provide, at least in the here 
used quenched approximation, a sensible definition of the 
neutral pion mass. This will allow us to test the scaling behaviour of
the mass splitting in Wilson twisted mass QCD.

In the Osterwalder-Seiler (OS) action
\cite{Osterwalder:1977pc,Frezzotti:2004wz} the $\tau_3$ matrix 
of eq.~(\ref{tmaction}) 
is replaced by a unit matrix in flavour space. With such an action 
there is no flavour breaking and the difference between neutral and charged 
pion masses vanishes, $m_{\pi^+}=m_{\pi^0}$.
For the OS action, 
\begin{equation}
C_{\pi^{+}}^\mathrm{OS}(x_0)= C_{\pi^{0}}^\mathrm{OS}(x_0)  
\label{oscorr}
\end{equation}
where no 
disconnected piece appears. 
In contrast, for the twisted mass action we have 
\begin{equation}
 C_{\pi^{0}}^\mathrm{tm}(x_0) = 
(C_{\pi^{0}}^\mathrm{tm}(x_0))_\mathrm{con} +
(C_{\pi^{0}}^\mathrm{tm}(x_0))_\mathrm{disc}\; .
\label{tmcorr}
\end{equation}
Now, the crucial observation is that    
\begin{equation}
 C_{\pi^{0}}^\mathrm{OS}(x_0) = 
(C_{\pi^{0}}^\mathrm{tm}(x_0))_\mathrm{con}\; .
\label{tmoscorr}
\end{equation}

Thus $(C_{\pi^{0}}^\mathrm{tm}(x_0))_\mathrm{con}$ can be interpreted 
as the correlation function of a local operator and has therefore
a standard transfer matrix decomposition. 
In particular, the exponential decay of 
$(C_{\pi^{0}}^\mathrm{tm}(x_0))_\mathrm{con}$ will allow us to 
extract the neutral pion mass and, since we can neglect the 
disconnected piece, the correlation function can be evaluated with 
good precision. Let us, nevertheless, 
emphasise that it is the main goal of this investigation to check 
whether the mass splitting shows the expected $O(a^2)$ lattice artifacts
and to estimate the size of the flavour breaking effects. 

In table~\ref{table:parameters} we give the simulation parameters and the 
statistics of our quenched runs. We give in 
table~\ref{table:kappac} the values for the critical hopping parameter 
obtained from the pion and the PCAC definitions of the critical mass. 
Finally, we show in table~\ref{table:piplus}
our results for the charged pion mass and in table~\ref{table:pineutral}
for the neutral pion mass. 
Note that the data in tables~\ref{table:kappac} and \ref{table:piplus} 
are the same as in  
ref.~\cite{Jansen:2005kk} and are only given here for completeness. 


\begin{table}[!t]
\begin{center}
\begin{tabular}{|c||c|c|c|c|c|c|}
\hline
\hline
$\beta$  &  5.85 & 6.00  & 6.20  \\   
$a$ (fm) & 0.123 & 0.093 & 0.068 \\
 $r_0/a$ & 4.067 & 5.368 & 7.360 \\
 $L/a$   & 16    & 16    & 24    \\
 $T/a$   & 32    & 32    & 48    \\
\hline
\hline
&\multicolumn{3}{|c|}{pion definition of $\kappa_c$}\\
\hline
$N_{\rm meas}$ & 378 & 387 & 260 \\
\hline
$\mu_1 a$& 0.0050 & 0.0038 & 0.0028 \\
$\mu_2 a$& 0.0100 & 0.0076 & 0.0055 \\
$\mu_3 a$& 0.0200 & 0.0151 & 0.0111 \\
$\mu_4 a$& 0.0400 & 0.0302 & 0.0221 \\
$\mu_5 a$& 0.0600 & 0.0454 & 0.0332 \\
$\mu_6 a$& 0.0800 & 0.0605 & 0.0442 \\
$\mu_7 a$& 0.1000 & 0.0756 & 0.0553 \\
\hline                                                     
\hline                                                     
&\multicolumn{3}{|c|}{PCAC definition of $\kappa_c$}\\
\hline
$N_{\rm meas}$ & 500 & 400 & 300 \\
\hline
$\mu_1 a$& 0.0050 & 0.0038 & 0.0028 \\
$\mu_2 a$& 0.0100 & 0.0076 & 0.0055 \\
$\mu_3 a$& 0.0200 & 0.0151 & 0.0111 \\
$\mu_4 a$& 0.0400 & 0.0302 & 0.0221 \\
$\mu_5 a$& 0.0600 & 0.0454 & 0.0332 \\
$\mu_6 a$& 0.0800 & 0.0605 & 0.0442 \\
$\mu_7 a$& 0.1000 & 0.0756 & 0.0553 \\
\hline
\hline
\end{tabular}
\end{center}
\caption{\it Simulation parameters}
\label{table:parameters}
\end{table}

\begin{table}[!t]
\begin{center}
\begin{tabular}{|c||c|c|}
\hline
\hline
$\beta$ & pion $\kappa_c$  & PCAC $\kappa_c$ \\
\hline
\hline
5.85 & 0.161662(17) &  0.162379(93)  \\ 
6.0  & 0.156911(35) &  0.157409(72)  \\
6.2  & 0.153199(16) &  0.153447(32)  \\
\hline
\hline
\end{tabular}
\end{center}
\caption{\it Critical values of the hopping parameters obtained from the
vanishing of the pseudoscalar meson mass (pion $\kappa_c$) and from the
vanishing of the PCAC mass (PCAC $\kappa_c$).}
\label{table:kappac}
\end{table}


\begin{table}[!t]
\begin{center}
\begin{tabular}{|c||c|c|c|c|c|c|}
\hline
\hline
$\beta$ &  5.85 & 6.00 & 6.20 \\   
\hline
\hline
&\multicolumn{3}{|c|}{$m_{\pi^+} a$ (pion $\kappa_c$)}\\
\hline
$\mu_1 a$& 0.1682(26) &  0.1385(66) &  0.1004(27) \\
$\mu_2 a$& 0.2256(22) &  0.1764(42) &  0.1298(23) \\
$\mu_3 a$& 0.3122(19) &  0.2373(32) &  0.1768(17) \\
$\mu_4 a$& 0.4452(14) &  0.3335(22) &  0.2463(15) \\
$\mu_5 a$& 0.5535(12) &  0.4134(17) &  0.3037(13) \\
$\mu_6 a$& 0.6488(13) &  0.4839(16) &  0.3546(12) \\
$\mu_7 a$& 0.7358(12) &  0.5491(14) &  0.4021(11) \\
\hline                                                     
\hline                                                     
&\multicolumn{3}{|c|}{$m_{\pi^+} a$ (PCAC $\kappa_c$)}\\
\hline
$\mu_1 a$& 0.1640(23) &  0.1217(66) &  0.0934(24) \\
$\mu_2 a$& 0.2289(17) &  0.1708(50) &  0.1276(21) \\
$\mu_3 a$& 0.3232(13) &  0.2396(33) &  0.1779(18) \\
$\mu_4 a$& 0.4606(11) &  0.3403(22) &  0.2492(13) \\
$\mu_5 a$& 0.5701(10) &  0.4214(17) &  0.3071(12) \\
$\mu_6 a$& 0.6658(9)  &  0.4925(14) &  0.3588(10) \\
$\mu_7 a$& 0.7530(9)  &  0.5579(14) &  0.4062(9)  \\
\hline
\hline
\end{tabular}
\end{center}
\caption{\it Pseudoscalar meson masses $m_{\pi^+} a$ for all simulation points.}
\label{table:piplus}
\end{table}


\begin{table}[!t]
\begin{center}
\begin{tabular}{|c||c|c|c|}
\hline
\hline
$\beta$ &   5.85 & 6.00 &  6.20  \\   
\hline
\hline
&\multicolumn{3}{|c|}{$m_{\pi^0} a$ (pion $\kappa_c$)}\\
\hline
$\mu_1 a$&   0.245(12) &  0.1858(84)&   0.1150(47) \\
$\mu_2 a$&   0.2989(63)&  0.2202(43)&   0.1419(27) \\
$\mu_3 a$&   0.3759(58)&  0.2751(28)&   0.1887(16) \\
$\mu_4 a$&   0.4974(31)&  0.3621(20)&   0.2563(12) \\
$\mu_5 a$&   0.5983(25)&  0.4371(17)&   0.3119(12) \\
$\mu_6 a$&   0.6887(23)&  0.5047(14)&   0.3615(11) \\
$\mu_7 a$&   0.7726(23)&  0.5670(14)&   0.4082(10) \\
\hline                                                     
\hline                                                     
&\multicolumn{3}{|c|}{$m_{\pi^0} a$ (PCAC $\kappa_c$)}\\
\hline
$\mu_1 a$&  0.319(14)  &  0.2183(92) &  0.123(10)  \\
$\mu_2 a$&  0.3510(79) &  0.2458(51) &  0.1537(50) \\
$\mu_3 a$&  0.4137(44) &  0.2929(29) &  0.1970(22) \\
$\mu_4 a$&  0.5276(24) &  0.3746(20) &  0.2615(14) \\
$\mu_5 a$&  0.6255(22) &  0.4478(15) &  0.3163(12) \\
$\mu_6 a$&  0.7143(19) &  0.5150(14) &  0.3655(12) \\
$\mu_7 a$&  0.7969(18) &  0.5782(13) &  0.4120(11) \\
\hline
\hline
\end{tabular}
\end{center}
\caption{\it Neutral pion masses $m_{\pi^0} a$ coming from the 
connected correlators for all simulation points.}
\label{table:pineutral}
\end{table}

\begin{figure}[htb]
\begin{center}
\epsfig{file=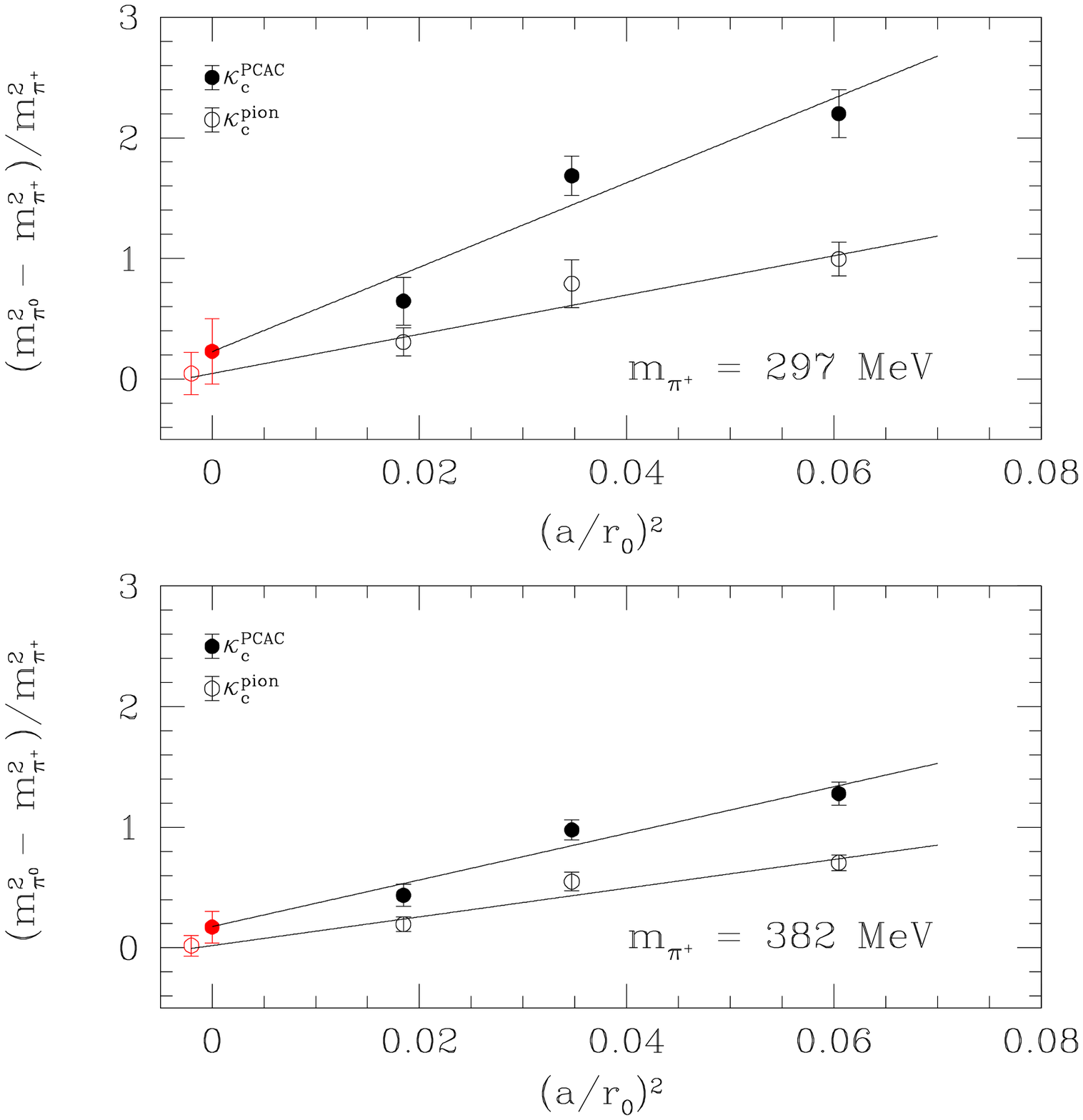,width=15.0cm}
\end{center}
 \caption{Relative pion mass difference
as function of $(a/r_0)^2$ at two fixed values of the charged pion mass
employing the pion and the PCAC  definitions of the critical point.
 \label{fig:piondiff}}
\end{figure}

In fig.~\ref{fig:piondiff} we show the relative difference 
$(m_{\pi^{0}}-m_{\pi^{+}})/m_{\pi^{+}}$ as a function of $(a/r_0)^2$, where 
$r_0=0.5$ fm is the Sommer parameter \cite{Sommer:1993ce}. 
For the plot we use data 
from the pion definition (open circles) and the PCAC definition 
(filled circles) fixing the pion mass to be $297$ MeV and $382$ MeV. 

Fig.~\ref{fig:piondiff} demonstrates that the scaling of the  mass
splitting is consistent with an $a^2$ behaviour as expected.  On the
other hand, as we already discussed in the previous section,  the
flavour breaking effects turn out to be rather large. For the case of
the conventional twisted theory, with disconnected  contributions to the
$\pi^0$ included, the errors are sufficiently large that a  precise
estimate of the scaling behaviour of the mass difference cannot be
obtained.  However, we have shown that the size of the flavour breaking
effects is reduced when the disconnected piece of the correlation
function is included. 

Another somewhat  surprising observation is that the flavour breaking
effect is larger  for the PCAC definition of
$\kappa_\mathrm{c}$ than for the  pion definition of
$\kappa_\mathrm{c}$.  We interpret this phenomenon in the following
way: for Wilson twisted mass  fermions at non-vanishing twist angle it
is not possible to conserve {\em simultaneously} parity and flavour 
symmetry.  Using the  PCAC definition of $\kappa_\mathrm{c}$, parity
is restored as well as  possible, leading to a large violation of
flavour symmetry. When using  the pion definition of
$\kappa_\mathrm{c}$ one looses somewhat  on the parity violation but
gains on the flavour symmetry.  This interpretation follows from an 
analytical study in the framework of Wilson chiral perturbation 
theory~\cite{Scorzato:2004da,Sharpe:2004ny}. 
With $\omega$  the twist angle, it is shown in these references that the
mass splitting is proportional to $\mathrm{sin}(\omega)$  indicating a
maximal flavour symmetry violation at maximal twist angle.

\section{Conclusion}

In this paper we have studied the flavour breaking effects of Wilson 
twisted mass lattice QCD in the quenched approximation. We focussed on
the  example of the mass splitting of the neutral to charged pion. 
While the charged pion mass has been computed with standard methods,  we
evaluated the neutral pion correlator by means of a stochastic  source
method. 
The absolute errors from the stochastic evaluation of disconnected 
contributions do not decrease with increasing $t$, unlike for the
connected case.
Since the pions we use
are relatively light and the disconnected contribution to  the $\pi^0$
propagators is relatively large, we are in a very favourable situation 
to evaluate disconnected contributions. 
 We found that about $24$ stochastic sources are  sufficient to keep the
fluctuations from the stochastic noise  under control and at a
comparable level to the error from the  gauge field average. This allows
an evaluation  of the disconnected contribution out to large $t$.  The
disconnected contribution to the $\pi^0$ is indeed  large and reduces
the apparent flavour violation substantially. 
We remark that the mass splitting $\Delta m=m_{\pi^0} - m_{\pi^+}$ comes out to 
be positive which is consistent with an Aoki phase scenario in the 
quenched approximation of lattice QCD. 
In addition, using the
Osterwalder-Seiler interpretation  of the twisted mass neutral 
pion correlation function, we computed the
neutral pion mass using solely the connected piece of the neutral pion
correlator which gives a more accurate result, albeit for  a less
familiar theory.

Our results reveal that, for the situation of full twist as has been
realized in this work, the mass splitting and hence the flavour breaking
 effects vanish with a rate that is quadratic in the lattice spacing as 
expected and at  a level characterised by
$r_0^2 (m_{\pi^0}^2-m_{\pi^+}^2)=c(a/r_0)^2$ with $c \approx 10$. 
Note that without adding the disconnected piece, this value of $c$ 
can easily be twice bigger. On the other
hand, for a fixed value of the lattice spacing of,  say, $a=0.1$ fm, the
flavour breaking effects are substantial and are not negligible. 
Indeed, comparing to a quenched simulation for naive staggered fermions 
with Wilson gauge action~\cite{Ishizuka:1993mt}, one finds 
a similar size of the flavour splitting effects encountered
for the pion mass at a similar lattice spacing with a value of $c \approx 40$. 
It will be very interesting to see, whether for dynamical fermions the
value of $c$ would change. The only source of information is presently
provided by dynamical improved staggered fermions~\cite{Aubin:2004wf} which
give  a value of $c$ similar in magnitude to the value we found above. One
should keep in mind, however, the caveats of the improved staggered simulations
using the fourth root trick~\cite{Bunk:2004br,Kennedy:2004ae}.  

The inclusion of the disconnected piece in the analysis turned out
to be crucial to reduce the flavour breaking effects. 
 An interesting observation is that with the PCAC definition of the
critical point that  reduces parity violations as much as possible, the
connected contributions to the flavour breaking  effects are larger than
with a definition of the critical point that leaves some parity 
violations. We interpret this phenomenon as arising  since it is not
possible to simultaneously keep both effects small and that there is a
trade-off between both. 

The work presented here has been performed in the quenched approximation 
and it is the aim of the paper to obtain a first insight into the flavour
breaking effects in Wilson twisted mass QCD. It is clear that the quenched 
approximation, in particular in the flavour singlet aspects, has 
conceptual short comings. This we see clearly for example in the correlator 
of $a_0$ which turns negative. Nevertheless, we think that the quenched 
approximation has provided useful insight into the flavour breaking effects 
of Wilson twisted mass fermions. 
Of course, it will be very interesting, to investigate flavour singlet 
quantities also for full QCD in this setup. 

\vspace*{-1mm}
\section*{Acknowledgements}
 We thank R. Frezzotti for useful discussions. The computer centres at
NIC/DESY Zeuthen, NIC at Forschungszentrum J{\"u}lich and HLRN provided
the technical help and computer resources. We acknowledge computer time
provided by the ULgrid project of the University of Liverpool.   We
thank Robert Edwards~\cite{Edwards:2004sx} for helping us test the
twisted inverter used to compute the disconnected loops. This work was
supported by the DFG  Sonderforschungsbereich/Transregio SFB/TR9-03.

\bibliographystyle{JHEP}
\bibliography{singlet}

\end{document}